\documentclass[%
 reprint,
 amsmath,amssymb,
 aps,
]{revtex4-2}
\usepackage{graphicx}
\usepackage{dcolumn}
\usepackage{bm}
\usepackage{color, xcolor}
\usepackage{appendix}
\newcommand {\bnabla} {\mbox{\boldmath$\nabla$}}

\begin{document}

\title{Field theory of structured liquid dielectrics}

\author{Ralf Blossey}
\affiliation{University of Lille, Unit\'e de Glycobiologie Structurale et Fonctionnelle (UGSF) CNRS UMR8576, 59000 Lille, France}
\author{Rudolf Podgornik}
\email{podgornikrudolf@ucas.ac.cn}
\affiliation{School of Physical Sciences and Kavli Institute for Theoretical Sciences, University of Chinese Academy of Sciences, Beijing 100049, China and CAS Key Laboratory of Soft Matter Physics, Institute of Physics, Chinese Academy of Sciences, Beijing 100190, China \\
and  Wenzhou Institute of the University of Chinese Academy of Sciences, Wenzhou, Zhejiang 325000, China}
\altaffiliation{Also affiliated with: Department of Theoretical Physics, Jo\v zef Stefan Institute, SI-1000 Ljubljana, Slovenia and Department of Physics, Faculty of Mathematics and Physics, University of Ljubljana, SI-1000 Ljubljana, Slovenia}

\date{\today}

\begin{abstract}

We develop of a field-theoretic approach for the treatment of both the non-local and the non-linear response of structured liquid dielectrics. Our systems of interest are composed of dipolar solvent molecules and simple salt cations and anions. We describe them by two independent order parameters, the polarization field for the solvent and the charge density field for the ions, including and treating the non-electrostatic part of the interactions explicitly and consistently. We show how to derive functionals for the polarization and the electrostatic field of increasingly finer scales and solve the resulting mean-field saddle-point equations in the linear regime. We derive criteria for the character of their solutions that depend on the structural lengths and the polarity of the solvent. Our approach provides a systematic way to derive generalized polarization theories.  
\end{abstract}


\maketitle

\section{Introduction} 

Phenomenological functionals of polarization have featured prominently in the development of the electrostatic theory of structured dielectrics starting from the seminal contributions of Marcus in the 1950's on electron 
transfer theory based on local dielectric response theory \cite{Marcus1956a,Marcus1956b}. Aqueous dielectrics, {\sl de rigueur} in soft- and bio-matter systems \cite{Holm1989}, pose another challenge as the dielectric response of water is strongly non-local and cannot be exhaustively described by a local dielectric approximation \cite{Kornyshev1996}. This has been the driving force behind the phenomenological non-local dielectric function methodology developed extensively by Kornyshev and collaborators \cite{Kornyshev1978,Kornyshev1986,Kornyshev1996}, as well as the Landau-Ginzburg-type free energy functionals describing the non-local dielectric response in spatially confined dielectrics  \cite{Maggs2006,Mertz2007,Berthoumieux2019,Berthoumieux2021,Monet2021}, the two approaches being of course equivalent in the bulk but differing in systems with boundaries.  

While it would be of course desirable to have more microscopic theories, aqueous dielectrics have consistently eluded descriptions based on more detailed explicit molecular interactions. As a successful example of a microscopic non-local dielectric theory one can consider the statistical mechanical formulation of the non-local dielectric properties of ice by Onsager and Dupuis \cite{Onsager1959,Onsager1962}, later further elaborated and generalized by Gruen and Mar\v celja \cite{Gruen1983a, Gruen1983b}. Of all the methodologies applied to non-local dielectrics this is the first to be found in the literature that implements not only the non-local but also the non-linear dielectric response of a system. It leads to {\sl two coupled equations} for the electrostatic field and the dielectric polarization field, and in general cannot be reduced to either the non-local dielectric function methodology or the local Landau-Ginzburg polarization functional. 
The coupling between the polarization and electrostatic fields is provided by the structural interaction mediated by the Bjerrum structural defects in the ideal hydrogen bonded lattice of ice. In its absence the basic equations reduce to the standard dielectric response theory. Being based on the crystalline lattice and hydrogen-bonding defects there is no simple way to use Onsager-Dupuis theory straightforwardly for a 
liquid aqueous dielectric. 

An important area of application for the non-local dielectric response of structured dielectrics was and remains the theory of hydration forces \cite{Leikin1993,Parsegian2011} where the dielectric response of ice \cite{Gruen1981} as well as the non-local electrostatics of liquid water \cite{Belaya1986,Kornyshev1986} were applied to water confined between substrate surfaces with the propensity for local ordering of vicinal water molecules. This type of water structural interaction modeling has been applied to solvation of phospholipid membranes \cite{Marcelja1976}, DNA molecules \cite{Rau1984}, linear polysaccharides \cite{Rau1990} or to the solvation of proteins \cite{Hildebrandt2004,Hildebrandt2007}. More recent approaches extend the ideas of non-local dielectric response of confined water \cite{Fumagalli2018} to order parameter description beyond the polarization \cite{Papadopoulou2021}, which are able to describe additional features of confined water structuring  \cite{Kanduc2014, Schlaich2016}. The order parameter description can be arguably viewed as a generalization of the Landau-Ginzburg-type free  energy functionals beyond the polarization formulation \cite{Mitlin1993}, describing additional degrees of freedom pertinent to the complicated hydrogen-bond phenomenology \cite{Varghese2019}. The hydration force theory also sheds light on  the fact that part of the water polarization is purely non-electrostatic in nature, being a separate field, independent of the imposed electrostatic field. This is in fact one of the most important conclusion of the Onsager-Dupuis theory of the dielectric response of ice  \cite{Onsager1962,Gruen1983a}. 

Motivated by the Onsager-Dupuis approach we develop a systematic field theory for the {\sl non-local as well as  non-linear dielectric response} of structured liquid dielectrics based upon the electrostatic field and dielectric polarization as 
{\sl separate order parameters}. From the outset, our approach decouples electrostatic and non-electrostatic structural contributions to these order parameters, allowing for the consistent inclusion of the effects that could not be implemented in previous approaches. On the linear level our theory reduces to two equations for the electrostatic and polarization potentials, which are coupled by the non-electrostatic interaction potentials and assume a form completely consistent with the linearized Onsager-Dupuis theory of ice. We explicitly discuss three examples of linear models differing by the degree of structural detailed included, and show how they relate to previously studied simplified models. 

\section{Order parameters and microscopic interactions}
 
The systems we consider are composed of three components: rigid rod-like solvent with a dipolar moment (N), positive simple salt ions (+) and negative simple salt ions (-). 
We first introduce two separate order parameters: charge density and polarization (dipolar charge density). For the solvent (N) component the order parameter is the  {\sl polarization field} 
defined as
\begin{equation}
\hat{\cal \bf P}({\bf x}) \equiv \sum_{(N)}  p\, {\bf n} ~ \delta\left( {\bf x} - {\bf x}_n \right)
\label{cole2}
\end{equation}
where $p$ is the dipolar moment of the molecule, so that the bound charge density field of the solvent material is
\begin{equation}
\hat{\rho}_{(N)}({\bf x}) \equiv  p \sum_{N} {\bf n}\cdot\bnabla~ \delta\left( {\bf x} - {\bf x}_n \right) = \bnabla\!\cdot\hat{\cal \bf P}({\bf x}),
\end{equation}
with dipoles considered to be point-like. For the simple salt component we can define the total charge density $\hat{\rho}_{(\pm)}({\bf x})$ as the sum of {\sl cation} and {\sl  anion density fields}
\begin{equation}
\hat{\rho}_{(\pm)}({\bf x}) \equiv e \sum_{N_+} \delta({\bf x} - {\bf x}_n)   - e \sum_{N_-} \delta({\bf x} - {\bf x}_n),
\end{equation}
where $e$ is the elementary charge of the salt ions, $N$ is the number of dipoles while $N_{\pm}$ are the numbers of positive and negative ions, respectively. The {\sl total charge density field} is then  
\begin{equation}
 \hat{\rho}({\bf x}) = \hat{\rho}_{(\pm)}({\bf x}) + \bnabla\!\cdot\hat{\cal \bf P}({\bf x}).
 \label{cole3}
\end{equation}

We will now first write the interactions in terms of these microscopic order parameters and then in the form of macroscopic collective coordinates, in exactly the same way as this is done for standard Coulomb fluids and/or for nematic Coulomb fluids \cite{Podgornik2021}. 
The interactions in the system have the following components: the non-electrostatic interactions in the harmonic approximation
\begin{equation}
{{\cal H}_{SR}} =  {\textstyle\frac12} \int\!\!\int_V d{\bf x} d{\bf x}' ~ \hat{\cal \bf P}_{i}({\bf x})\tilde u_{ij}({\bf x} -{\bf x}') \hat{\cal \bf P}_{j}({\bf x}'), 
\label{hamilt0}
\end{equation}
where we assumed that the non-electrostatic tensorial part $\tilde u_{ij}({\bf x} -{\bf x}') $ is a {\it short-range, non-electrostatic potential}; the electrostatic interactions are given by the standard Coulomb form
\begin{equation}
{{\cal H}_C} = {\textstyle\frac12} \int\!\!\int_V d{\bf x} d{\bf x}' ~ {\hat\rho}({\bf x})u({\bf x}\!-\!{\bf x}') {\hat\rho}({\bf x}'),
\label{hamilt1}
\end{equation}
with the proviso that we take the Coulomb potential with the non-configurational dielectric constant, or the high frequency dielectric constant $\varepsilon_{\infty}$ as in the Onsager-Dupuis model, {\sl i.e.}
\begin{equation}
u({\bf x} -{\bf x}') = ({1}/{4\pi {\textstyle \varepsilon_{\infty}\varepsilon_0})~ \vert {\bf x} -{\bf x}' \vert}^{-1}\, ,
\label{Coulomb}
\end{equation}
where $\varepsilon_{\infty}$  is associated with all the relaxation mechanisms apart from the dipolar one; further, we allow for non-electrostatic interactions due to the hydration shell of the ions, which generalizes the role played by the ion-bound Bjerrum defects in the Onsager-Dupuis theory of ice \cite{Gruen1981}.
This effect corresponds to the coupling between the ion density and $\hat{\rho}_{(\pm)}({\bf x})$ and $\hat{\cal \bf P}_{j}({\bf x})$. To the lowest order this coupling can be written via the hydration energy of the form (for details see Appendix B)
\begin{equation}
{{\cal H}_{HY}} =  \int\!\!\int_V d{\bf x} d{\bf x}' ~  \hat{\rho}({\bf x}) \tilde{u}({\bf x} -{\bf x}') \bnabla\!\cdot\hat{\cal \bf P}({\bf x}'),
\label{hamilt3}
\end{equation}
where the potential $\tilde u({\bf x} -{\bf x}')$ is again a {\it short-range, non-electrostatic potential} for the hydration shell interactions. 
Here we assume for simplicity that the polarization in the hydration shell of anions and cations is - apart from its direction - the same. 

The total interaction energy functional is thus composed of the standard electrostatic Coulomb interaction between all the charges in the system, 
but in a medium with high frequency dielectric constant, and the non-electrostatic short-range polarization interactions and ion polarization interactions, and is written
as
\begin{equation}
\label{Hamiltonian}
{\cal H} =  {\cal H}_{C}[{\hat\rho}({\bf x})] + {{\cal H}_{SR}}[{\hat{\bf P}({\bf x})] + {{\cal H}_{HY}}[{\hat\rho}({\bf x}), {\hat{\bf P}}({\bf x})],}
\end{equation}
so that the interaction Hamiltonian is ${\cal H} \equiv {\cal H}[{\hat\rho}({\bf x}), {\hat{\bf P}}({\bf x})]$. We note that the interaction Hamiltonian need not be quadratic and our approach can easily incorporate higher-order interactions.

\section{Collective description and field theory}

From here we derive the partition function in terms of two order parameters and two auxiliary fields (for details, see Appendix A) as
\begin{eqnarray}
{\Xi} &\equiv& \int {\cal D}[{\cal \bf P}_{i}({\bf x})]{\cal D}[{\cal E}_i({\bf x})] ~ \int {\cal D}[\rho({\bf x})]{\cal D}[\phi({\bf x})]~ \times \nonumber\\
&& e^{-\beta {\cal H}[ {\cal \bf P}_i({\bf x}), \rho({\bf x}); {\cal E}_i({\bf x}), \phi({\bf x})]}.
\end{eqnarray}
This expression differs from the standard Coulomb fluid field-theoretical representation of the partition function \cite{Edwards1962,Podgornik1989} by the presence of two additional independent fields ${\cal \bf P}_{i}({\bf x})$ and ${\cal E}_i({\bf x})$. 
In what follows we will not analyze the general features of the partition function, but will limit ourselves immediately to its {\sl saddle-point} evaluation, which is taken with respect to the auxiliary fields ${\cal E}_i({\bf x}) \longrightarrow i{\cal E}^*_i({\bf x})$ and $\phi({\bf x}) \longrightarrow i \phi^*({\bf x}) $. The partition function is then approximated by 
$ - \ln{{\Xi}} \longrightarrow  \beta {\cal H}[{\cal \bf P}_i({\bf x}), \rho({\bf x}); i{\cal E}^*_i({\bf x}), i\phi^*({\bf x})] 
$
where the effective-field Hamiltonian is obtained from the partial trace over coordinate and orientation degrees of freedom for the ions and dipoles in the partition function interacting {\sl via} the interaction energy functional Eq. (\ref{Hamiltonian}). It can be obtained in the form 
\begin{widetext}
\begin{equation}
\beta {\cal H}[{\cal \bf P}_i({\bf x}), \rho({\bf x}); i{\cal E}^*_i({\bf x}), i\phi^*({\bf x})] = 
\label{maineq1}
{\cal H}[{\cal \bf P}_i({\bf x}), \rho({\bf x})] - \int_V d{\bf x} ~{\cal \bf P}_{i}({\bf x}) {\cal E}^*_{i}({\bf x}) - \int_V d{\bf x} ~\rho({\bf x})\phi^*({\bf x}) -  \int_V d^3{\bf x} ~v^*\left({\cal E}^*_i({\bf x}), \phi^*({\bf x})\right), 
\end{equation}
\end{widetext}
where the first term is the interaction Hamiltonian, Eq. (\ref{Hamiltonian}).

The expression for the effective-field interaction potential $v^*({\cal E}_i({\bf x}), \phi^*({\bf x})) = v(i{\cal E}^*_i({\bf x}), i \phi^*({\bf x}))$ depends crucially on the model used for the dipolar fluid.  We specifically obtain its form for two models: the case of a {\sl dipolar  model (DM)} is obtained without any constraints on the summation \cite{Levy2013}, while in the case of the {\sl dipolar Langevin model (DLM)} one assumes an underlying lattice of spacing $a$ (roughly equal to the molecular size) \cite{Adar2018}, imposing the condition that each site of the lattice is occupied by only one of the species (incompressibility condition). 

For the DM-model the effective-field interaction potential can be written explicitly (see Appendix A) as 
\begin{eqnarray}
v_{DM}^*({\cal E}^*_i({\bf x}), \phi^*({\bf x})) = \lambda ~\Big( \frac{\sinh{u}}{u}+ 2 \tilde\lambda_s  \cosh{w}\Big) 
\label{Xi-2}
\end{eqnarray}
where $u = \beta p ~\vert {\cal E}^*({\bf x})-\bnabla \phi^*({\bf x}) \vert$ and $w = \beta e \phi^*({\bf x})$. 
The DM model implies no constraints on the local density of either ions or dipoles, 
and thus exhibits a form of an ideal van't Hoff gas of ions and dipoles \cite{Abrashkin2007,Levy2013}. 
For the DLM model we have 
$ v^*_{DLM} = \frac{1}{a^3}\ln\Big(v^*_{DM}/\lambda\Big)\,$. 

An important new feature of both models is the dependence of the effective field interaction potential on the field difference ${\cal E}^*({\bf x})-\bnabla \phi^*({\bf x})$.
The auxiliary field ${\cal E}^*({\bf x})$ here plays the role of a coupling field that quantifies the degree of non-electrostatic coupling between polarization and electrostatic fields and is a {\sl distinct feature} of our approach. 

\section{Mean-field (saddle-point) equations}
 
The saddle-point equations can be obtained for all the variables, {\sl i.e.} the order parameters as well as the auxiliary fields. For the two order parameters we obtain 
\begin{eqnarray}
\label{pol11}
{\cal E}^*_{i}({\bf x}) &=&  - \int_V d{\bf x}' \tilde u_{ij}({\bf x} -{\bf x}') {\cal \bf P}_{j}({\bf x}')  \nonumber \\
& & ~~~~~~~~ +  \bnabla_i\int_V d{\bf x}' ~\tilde u({\bf x} -{\bf x}') {\rho}({\bf x}'), \nonumber\\ 
\label{pol12}
\phi^*({\bf x}) &=&  \int_V d{\bf x}' u({\bf x} -{\bf x}') {\rho}({\bf x}') \nonumber\\
& &  ~~~~~~~~ + \int_V d{\bf x}' \tilde u({\bf x} -{\bf x}') \bnabla \cdot{{\cal \bf P}}({\bf x}'). 
\end{eqnarray}
Of the two equations the first one contains only non-electrostatic contributions, while the second one contains combined electrostatic - non-electrostatic contributions. 

After observing that the effective field interaction potential  $v^*[{\cal E}^*_i({\bf x}), \phi^*({\bf x})] = v^*[u , w]$, 
the two coupled saddle-point equations for the auxiliary fields can be written as 
\begin{eqnarray}
{\cal \bf P}({\bf x}) &=& - 
(\beta p)^2 \left(\frac{1}{u}\frac{\partial v^*}{\partial u}\right) \left({\cal E}^*({\bf x})-\bnabla \phi^*({\bf x})\right) \nonumber\\
&& {\rho}({\bf x}) - \bnabla \cdot {\cal \bf P} ({\bf x}) = - (\beta e) \frac{\partial v^*}{\partial w}.
\label{nonlineeq}
\end{eqnarray}
Eqs. (\ref{nonlineeq}) together with Eqs. (\ref{pol11}) and (\ref{pol12}) constitute the mean-field description of the structured dielectric. 

In order to fully formulate our field equations we need to define the structural potentials. For this we expand the non-electrostatic potential in terms of its gradients and consider explicitly the local, second-order and fourth-order forms 
\begin{widetext}
\begin{eqnarray}
\tilde u_{ij}({\bf x} -{\bf x}') = \tilde u_P(0) \Big( \delta_{ij} \delta({\bf x} -{\bf x}') + \xi^2~\bnabla'_j\bnabla_i \delta({\bf x} -{\bf x}') + \zeta^4 \bnabla'_k\bnabla'_j\bnabla_k\bnabla_i \delta({\bf x} -{\bf x}') \Big)+  \dots
\label{shortrange1}
\end{eqnarray}
\end{widetext}
where $\tilde u_P(0), \xi$ and $\zeta$ are the material constants in the constitutive relations. The length $\xi$ corresponds to the solvent particle correlation length 
and $\zeta$ to the solvent molecular size. 

Eq. (\ref{nonlineeq}) for ${\bf P}({\bf x})$ can be expressed for the DM and DLM-models
as 
\begin{equation}
    {\bf P}({\bf x}) = - \lambda_0 (\beta p)^2 
    Q(u) \left({\cal E}^*({\bf x})-\bnabla \phi^*({\bf x})\right)\, ,
\end{equation}
where $\lambda_0 \equiv \lambda $ for the dipolar model and $\lambda_0 \equiv 1/a^3$ 
for the dipolar Langevin model. Writing 
\begin{equation}
    {\cal E}^*({\bf x}) = u_P(0){\bf P}({\bf x}) - {\cal F}(\bnabla \cdot {\bf P}({\bf x})) 
\end{equation}
we can express the polarization field as
\begin{equation}
    {\bf P}({\bf x}) = \frac{\lambda_0 (\beta p)^2 Q(u)}{1 + \lambda_0 u_P(0)(\beta p)^2 Q(u)}(\bnabla \phi^*({\bf x}) + {\cal F}(\bnabla \cdot {\bf P}))\, .  
\end{equation}
The function $Q(u)$ for the DM-model can be written in the form
\begin{equation}
     Q(u) = {\cal G}(u) \equiv \frac{1}{u}\frac{d}{du}\left(\frac{\sinh(u)}{u}\right)
\end{equation}
while for the DLM-model it reads as
\begin{equation} 
 Q(u) = {\cal H}(u) \equiv \frac{u}{\sinh(u)}{\cal G}(u)\, .
\end{equation}

The highly nonlinear dependence of the field equations on the polarization field and the electrostatic field via the function $Q(u)$ is illustrated in Figure 1 for the DM-model. It shows the function 
\begin{equation}
   {\cal G}^*(u) = \frac{\lambda (\beta p)^2 G(u)}{1 + \lambda u_P(0)(\beta p)^2 G(u)}
   \label{G(u)}
\end{equation}
together with the lowest-order nonlinear approximation from its Taylor-expansion around its minimum. 
 \begin{figure}[t] 
\begin{center}
\includegraphics[width=9cm]{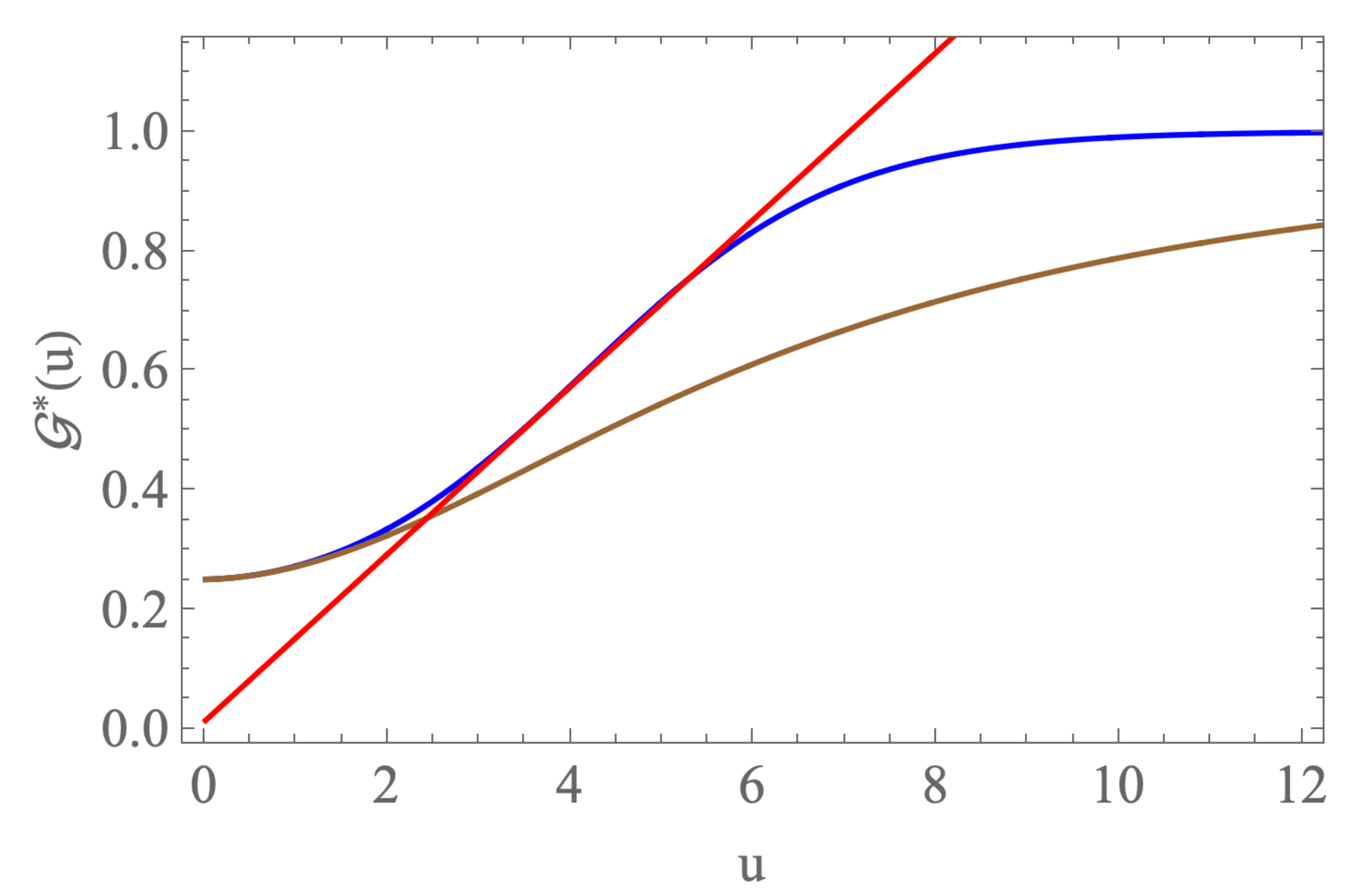}
\end{center}
\caption{${\cal G}^*(u)$ in the DM model (blue line) and the lowest-order nonlinear approximation to ${\cal G}^*(u)$ in an expansion around the minimum (brown line); linear approximation to the central region of the function (red line).}    
\label{Figure1}
\end{figure} 
${\cal G}^*(u)$ saturates at both small and large arguments and linearly interpolates between the two limiting values, see Figure 1.

\section{Linear model equations}

In view of the complexity of the fully nonlinear equations, we restrict our calculations in
the following to the linear case, identical for the DM- and DLM-models, corresponding to the lowest-order term in Eq. \eqref{G(u)}.
Truncating the non-electrostatic structural interaction potentials Eq. (\ref{shortrange1}) after the second-order and fourth-order derivative terms,
respectively, we refer to the resulting expressions as {\sl Models 1} and {\sl 2}. As the energy Eq. (\ref{hamilt3}) is already first order in the derivatives, the simplest possible local from for the ion density-polarization coupling is
\begin{equation}
\tilde u({\bf x} -{\bf x}') = \alpha ~\delta({\bf x} -{\bf x}') +  \dots 
\label{alphaform}
\end{equation} 
where $\alpha$ can be interpreted as the strength of the hydration shell of the ions. Considering an ion with effective radius $a$, charge $e$ and polarization of the hydration shell $P_n$ with a hydration (free) energy  ${\cal F}_{HY}$, 
we are led to identify $\alpha = {\textstyle\frac13} {\cal F}_{HY} a/(P_n e)$. Including the ion density-polarization non-electrostatic coupling given to the lowest order, we term this expression {\sl Model 3}. We now formulate the equations of these models explicitly.

{\it Model 1.} Stopping at the second-order derivative term in Eq. (\ref{shortrange1}), we have from Eqs. (\ref{pol11}) for the structural part the relation
\begin{equation}
{\cal E}^*({\bf x}) = u_P(0) \Big( {\bf P}({\bf x}) - \xi^2 \bnabla\left( \bnabla\cdot {\bf P}({\bf x})\right)\Big)\, .
\label{sp-nonlocal-linear}
\end{equation}
Employing the linearized saddle-point Eq. (\ref{nonlineeq}) with
\begin{eqnarray}
v^*[{\cal E}^*({\bf x}), \phi^*({\bf x}), \nabla\phi^*({\bf x}) ] = {\textstyle(\beta e)^2} \phi^*({\bf x})^2 + {\textstyle\frac{\lambda (\beta p)^2}{6}} u({\bf x})^2 \nonumber
\end{eqnarray}
we can eliminate the field ${\cal E}^*({\bf x})$ and find
\begin{equation}
u_P(0)\xi^2 \bnabla\left( \bnabla\cdot {\bf P}({\bf x})\right) - \Big(u_P(0) + \frac{3}{\lambda p^2}\Big) \nonumber
{\bf P}({\bf x}) = - \bnabla\phi^*({\bf x})\, 
\label{sp3} 
\end{equation}
\begin{eqnarray}
\bnabla \left( \varepsilon_{\infty}\varepsilon_0 \bnabla\phi^*({\bf x}) {+ {\cal \bf P}({\bf x})} \right)  =  {2(\beta e)^2 \lambda_s \phi^*({\bf x})}\, .
\label{sp-two-nonlocal-linear} 
\end{eqnarray}
\\
The coupled equations for $\phi^*({\bf x})$ and ${\bf P}({\bf x})$ correspond to the equations derived in Ref. \cite{Paillusson2010} with an opposite sign of the polarization field.

As in Refs. \cite{Gruen1983a, Paillusson2010} we introduce the {\sl polarization potential}
$\phi^{\dagger}({\bf x})$ which can be defined via 
\begin{equation}
{\bf P}({\bf x}) = (\varepsilon - \varepsilon_{\infty})\varepsilon_0 \bnabla\phi^{\dagger}({\bf x})\, ,
\label{polfield}
\end{equation}
making use of the identification 
\begin{equation}
     \varepsilon \varepsilon_0 \equiv \varepsilon_{\infty}\varepsilon_0 + \frac{{\textstyle\frac{1}{3}} \lambda p^2}{1 + u_P(0){\textstyle\frac{1}{3}} \lambda p^2} 
\label{def-epsilon2}     
\end{equation}
that relates the structural coupling $u_P(0)$ to the dielectric constants and the strength of the water dipole. 
Integrating the equation for ${\bf P} $, using the definition Eq. (\ref{def-epsilon2}) and defining
\begin{equation}
\widehat{\xi}^2 \equiv (\varepsilon - \varepsilon_{\infty})\varepsilon_0 u_P(0)\, \xi^2, 
\end{equation}
the mean-field equations for our Model 1 read as
\begin{eqnarray}
\bnabla^2\phi^{\dagger}({\bf x}) &=& {\widehat{\xi}^{-2}} \left( \phi^{\dagger}({\bf x}) - \phi^*({\bf x})\right),\\
 \bnabla^2\phi^*({\bf x})  &=& ~ {\textstyle\frac{\varepsilon}{\varepsilon_{\infty}}} \kappa_D^2 \phi^*({\bf x}) \nonumber \\ 
&& +\,\, {\widehat{\xi}^{-2}}\left( {\textstyle\frac{\varepsilon}{\varepsilon_{\infty}}} - 1 \right) \left( \phi^*({\bf x}) - \phi^{\dagger}({\bf x})\right)\, , \nonumber
\label{sp3c}
\end{eqnarray}
in which the inverse square of Debye length is defined by $\kappa_D^2 \equiv {2(\beta e)^2 \lambda_s}/{\varepsilon\varepsilon_0}$. Except for the sign of $\phi^{\dagger}$ these equations are the same as the Onsager-Dupuis equations \cite{Onsager1959}. 

{\it Model 2.} Keeping the terms up to the fourth order in derivatives in Eq. (\ref{shortrange1}) we find, introducing a second redefined length $ \widehat{\zeta}_0 $
\begin{equation}
\widehat{\zeta_0}^4 \equiv (\varepsilon - \varepsilon_{\infty})\varepsilon_0 u_P(0)\, \zeta^4 
\end{equation}
the set of equations 
\begin{eqnarray}
\widehat{\zeta_0}^4 \bnabla^2\bnabla^2\phi^{\dagger}({\bf x}) - {\widehat{\xi}^{2}}\bnabla^2\phi^{\dagger}({\bf x}) &=& \phi^*({\bf x}) - \phi^{\dagger}({\bf x}) ,  \nonumber\\
 \bnabla^2\phi^*({\bf x}) + \left({\textstyle\frac{\varepsilon}{\varepsilon_{\infty}}} - 1 \right) \bnabla^2 \phi^{\dagger}({\bf x})  &=& ~ {\textstyle\frac{\varepsilon}{\varepsilon_{\infty}}} \kappa_D^2 \phi^*({\bf x}),
\label{sp3c1}
\end{eqnarray}
which depend on three characteristic lengths, $1/\kappa_D, \xi$ and $\zeta_0$.

{\it Model 3.} The ion density-polarization coupling is incorporated {\sl via} the lowest order expansion Eq. (\ref{alphaform}) and gives rise to the following equations for the electrostatic and polarization potentials  
(for details on the derivation, see Appendix B) 
\begin{eqnarray}
\widehat{\zeta}^4 \bnabla^4 \phi^{\dagger}({\bf x})  - \widehat{\xi}^2 \bnabla^2 \phi^{\dagger}({\bf x})  + \widehat{\alpha}^2 \bnabla^2 \phi^*({\bf x}) =  \phi^*({\bf x}) - \phi^{\dagger}({\bf x}),\nonumber \\
\left(\textstyle{\frac{\varepsilon}{\varepsilon_{\infty}}} - 1\right)
[1  - \widehat{\alpha}^2 \bnabla^2] \bnabla^2 \phi^{\dagger} +  \bnabla^2 \phi^*  = \textstyle{\frac{\varepsilon}{\varepsilon_{\infty}}} \kappa_D^2 \phi^* \nonumber \\
\label{model3}
\end{eqnarray}
with the additional length $\widehat{\zeta}$  
\begin{equation}
\widehat{\zeta}^4 \equiv \left(u_P(0)\zeta^4 - \alpha^2  \varepsilon_{\infty}\varepsilon_0\right)  (\varepsilon-\varepsilon_{\infty}) \varepsilon_0 = \widehat{\zeta}^4_0 - \widehat{\alpha}^4\left({\textstyle\frac{\varepsilon}{\varepsilon_{\infty}}} - 1\right)
\\
\end{equation}
where the definition $ \widehat{\alpha}^2 \equiv \alpha \varepsilon_0 \varepsilon_{\infty} $ has been introduced. This model obviously depends on all four characteristic lengths that we consider in this model catalogue.

\section{Decoupling and solving the equations} 

Since we will consider one-dimensional geometries like a slit geometry $ -L \leq z \leq L$ or a single-plate geometry $0 \leq z < \infty$ we assume ${\bf x} \equiv (0,0,z)$ and the mean-field equations reduce to ordinary differential equations. The first step in solving these equations is then the decoupling of the electrostatic and polarization potentials, $\phi^*$ and $\phi^{\dagger}$, respectively. 

\subsection{Model 1} 

In the case of {\sl Model 1} the equations can be written in matrix form $\frac{d^2}{dz^2} \Phi_1(z) = {\cal M}_1 \Phi_1(z)$ with the introduction of the composite field $\Phi_1(z) \equiv (\phi^*(z),\phi^{\dagger}(z))$. The coupling matrix ${\cal M}_1$ is given by
\begin{eqnarray}
\hspace*{-0.5cm}{\cal M}_1 = \left(\begin{array}{cc}
    \frac{\varepsilon}{\varepsilon_{\infty}}\kappa_D^2 + \left(\frac{\varepsilon}{\varepsilon_{\infty}} - 1\right)\widehat{\xi}^{-2} &  - \left(\frac{\varepsilon}{\varepsilon_{\infty}}- 1\right)\widehat{\xi}^{-2} \\
    -\widehat{\xi}^{-2}      &  \widehat{\xi}^{-2} 
\end{array}\right)\, . \nonumber
\end{eqnarray}
Diagonalizing the matrix and rescaling  the eigenvalues with $\lambda \longrightarrow \lambda/\widehat{\xi}^2$ leads to a quadratic eigenvalue equation given by
\begin{equation}
\lambda^2 - \lambda~ {\textstyle\frac{\varepsilon}{\varepsilon_{\infty}}}\left( 1 + (\kappa_D\widehat{\xi})^2\right) + {\textstyle\frac{\varepsilon}{\varepsilon_{\infty}}}(\kappa_D\widehat{\xi})^2 = 0,
\end{equation}
where $\kappa^{\pm} = \sqrt{\lambda_{\pm}}$ are then the two inverse decay lengths corresponding to the two eigenvalues, which are both real and positive. They correspond to the decay lengths of the electrostatic and the polarization potentials, respectively, and coincide with the result of the linearized Onsager-Dupuis theory \cite{Gruen1983b,Paillusson2010}. 
The decay lengths are shown in Figure 2 as functions of $\kappa_D \hat\xi$ in accord with
earlier results, see \cite{Ruckenstein2002}.
\begin{figure}[tbh] 
\begin{center}
\includegraphics[width=8.9cm]{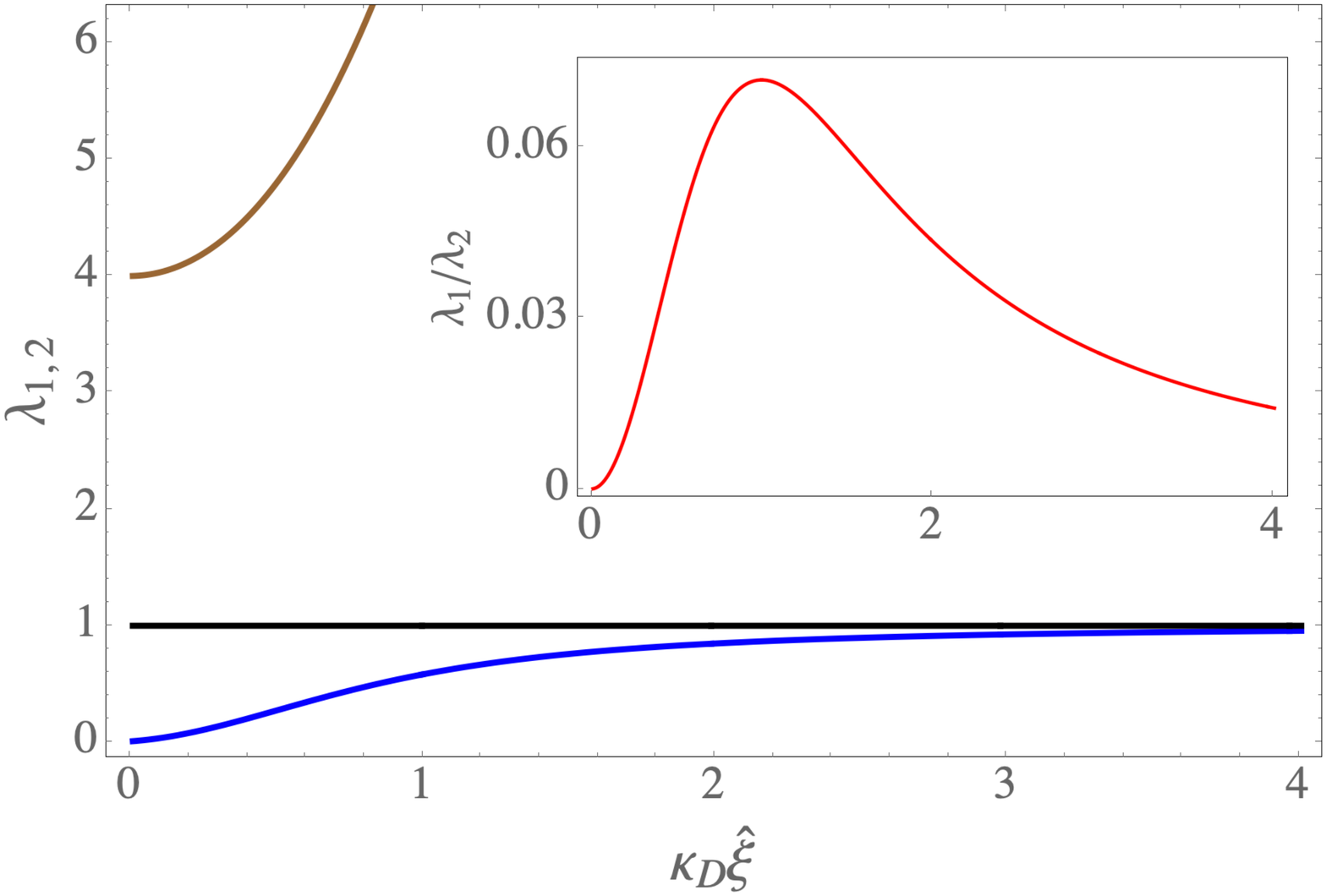}
\end{center}
\caption{The eigenvalues of the matrix ${\cal M}_1$, with $\lambda_1 \leq 1$, as a function of dimensionless parameter $\kappa_D \hat\xi$. The horizontal line corresponds to $\lambda=1$. Insert: the ratio of $\lambda_1$ and $\lambda_2$.}
\end{figure}

The key feature of our description, being the decoupling of the degrees of freedom of the electrostatic field and the polarization field, translates into the requirement of separate boundary conditions for both. In order to solve the linear model equations, we therefore need to specify the boundary conditions for $ -\bnabla \phi^* \cdot {\bf n}$ and ${\bf P}\cdot{\bf n}$ at the bounding surface(s) with normal vector ${\bf n}$. Alternatively we can assume that the bounding surfaces carry `polarization charges' $\sigma_P$
\begin{eqnarray}
\bnabla \cdot {\bf P}({\bf x}) = \rho_P({\bf x}) = \sigma_P \delta(z - z_{0}), 
\end{eqnarray}
as well as the standard electric charges $\sigma$ 
\begin{eqnarray}
\varepsilon_{\infty}\varepsilon_0\bnabla \cdot {\bf E}({\bf x}) = \rho({\bf x}) = \sigma \delta(z - z_{0}), 
\end{eqnarray}
where $z$ is the coordinate perpendicular to the boundary. 
Figure 3 shows a solution of {\sl Model 1} for the case of prescribed electric field
and polarization at the boundary of a slit of width $2L$, i.e., $z_{0} = \pm L$.
The solutions can be compared with the results derived in \cite{Ruckenstein2002}.
\begin{figure}[t] 
\begin{center}
\includegraphics[width=9cm]{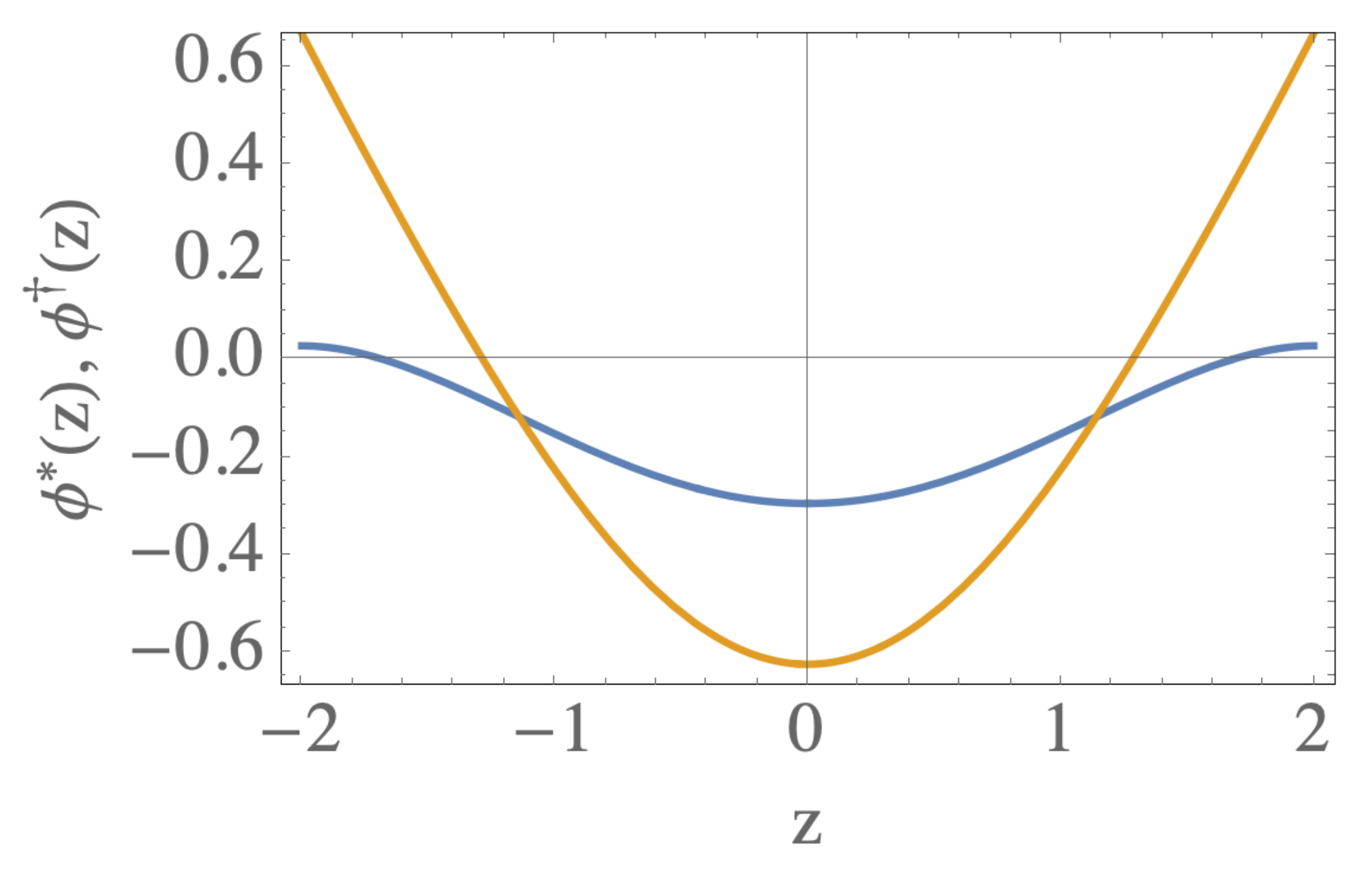}
\end{center}
\caption{Exemplary solutions of {\sl Model 1} for $\phi^*(z)$ (blue line) and $\phi^{\dagger}(z)$ (yellow line). The boundary conditions are: $(\phi^*)'(-L) = -E_0$,
$(\phi^*)'(L) = E_0$, $(\phi^{\dagger})'(-L) = (\phi^{\dagger})'(L) = 0$.
}    
\label{Figure3}
\end{figure} 

\subsection{Model 2}

For {\sl Model 2} (and later {\sl Model 3}), we can proceed in a similar manner. In this case we first have to reduce the order of the equations from four to two by introducing an auxiliary potential $\psi$ 
which in the one-dimensional geometry leads to 
\begin{equation}
 \frac{d^2\phi^{\dagger}(z)}{dz^2} = \frac{1}{\widehat{\xi}^{2}}\psi(z)\,. 
\end{equation}
In terms of the three-potential composite field $\Phi_2(z) \equiv (\phi^*(z),\phi^{\dagger}(z),\psi(z)) $ we now have the expression $ \frac{d^2}{dz^2} \Phi_2(z) = {\cal M}_2 \Phi_2(z) $, with the coupling matrix
\begin{eqnarray}
{\cal M}_2 = \left(\begin{array}{ccc}
    \frac{\varepsilon}{\varepsilon_{\infty}}\kappa_D^2  & 0 & -\left(\frac{\varepsilon}{\varepsilon_{\infty}} - 1\right)\widehat{\xi}^{-2}\\
    0        &  0 &   \widehat{\xi}^{-2}\\
    \frac{\widehat{\xi}^{2}}{\widehat{\zeta_0^{4}}} &  - \frac{\widehat{\xi}^2}{\widehat{\zeta_0}^4} & \frac{\widehat{\xi}^2}{\widehat{\zeta_0}^4}
\end{array}\right)\, .
\end{eqnarray}
The corresponding eigenvalue equations are thus of third order, depending on four parameters.  One can scale out one of them, remaining with the ratios. {\sl E.g.}, with the substitution $\lambda \longrightarrow \lambda/\widehat{\zeta_0}^2$ the characteristic polynomial is   given by an expression containing only the ratios of the model parameters 
\begin{eqnarray}
\hspace*{-1cm}     \lambda^3 &-& \lambda^2
    \left[\left({\textstyle\frac{\widehat{\xi}}{\widehat{\zeta_0}}}\right)^2 + {\textstyle\frac{\varepsilon}{\varepsilon_{\infty}}}(\kappa_D\widehat{\zeta_0})^2  \right] + \nonumber \\
&&+ ~\lambda 
    {\textstyle\frac{\varepsilon}{\varepsilon_{\infty}}}\left(1 + 
    (\kappa_D\widehat{\xi})^2 \right) 
    - 
     {\textstyle\frac{\varepsilon}{\varepsilon_{\infty}}}
    (\kappa_D \widehat{\zeta_0})^2 = 0\, .
    \label{eigenvalues1}
\end{eqnarray}
In order to see the character of the roots we need to look at the discriminant $\Delta$ of the cubic polynomial. Since both $(\kappa_D\widehat{\xi})^2 \ll 1$ and 
$ (\kappa_D\widehat{\zeta_0})^2 \ll 1$, the discriminant is approximately 
given by the expression
\begin{eqnarray}
&& \Delta = \left({\textstyle\frac{\varepsilon}{\varepsilon_{\infty}}}\right)^2  \left(\left({\textstyle\frac{\widehat{\xi}}{\widehat{\zeta_0}}}\right)^4 - 4{\textstyle\frac{\varepsilon}{\varepsilon_{\infty}}}\right) \\
&& - 4\left({\textstyle\frac{\widehat{\xi}}{\widehat{\zeta_0}}}\right)^2
\left(\textstyle\frac{\varepsilon}{\varepsilon_{\infty}}\right) 
\left(\left({\textstyle\frac{\widehat{\xi}}{\widehat{\zeta_0}}}\right)^4 - 4.5{\textstyle\frac{\varepsilon}{\varepsilon_{\infty}}}\right)
(\kappa_D\widehat{\zeta_0})^2 \nonumber \\
&& - 27\left({\textstyle\frac{\varepsilon}{\varepsilon_{\infty}}}\right)^2
(\kappa_D\widehat{\zeta_0})^4\, . \nonumber
\label{disc}
\end{eqnarray}
In the limit $\kappa_D\widehat{\zeta_0} \rightarrow 0$ one sees that the sign of
the discriminant depends on the difference
$
    \left({\textstyle\frac{\widehat{\xi}}{\widehat{\zeta_0}}}\right)^4 - 4{\textstyle\frac{\varepsilon}{\varepsilon_{\infty}}}
$
{\it i.e.} in a highly nonlinear way on the length-ratio $\xi/\zeta_0$ and the polarity of the solvent. 

Figure 4 shows the discriminant, Eq. (\ref{disc}), as a function of $\xi/\zeta_0$ and $\varepsilon/\varepsilon_{\infty}$. The function is cut by the zero level such that for positive $\Delta$ there are three real roots, while for 
negative $\Delta$ there is one real and two complex conjugate roots. The discriminant 
of {\sl Model 2} therefore signals a change in the character of the solutions from
exponential decay to damped oscillatory behaviour.
\begin{figure}[tbh] 
\begin{center}
\includegraphics[width=9cm]{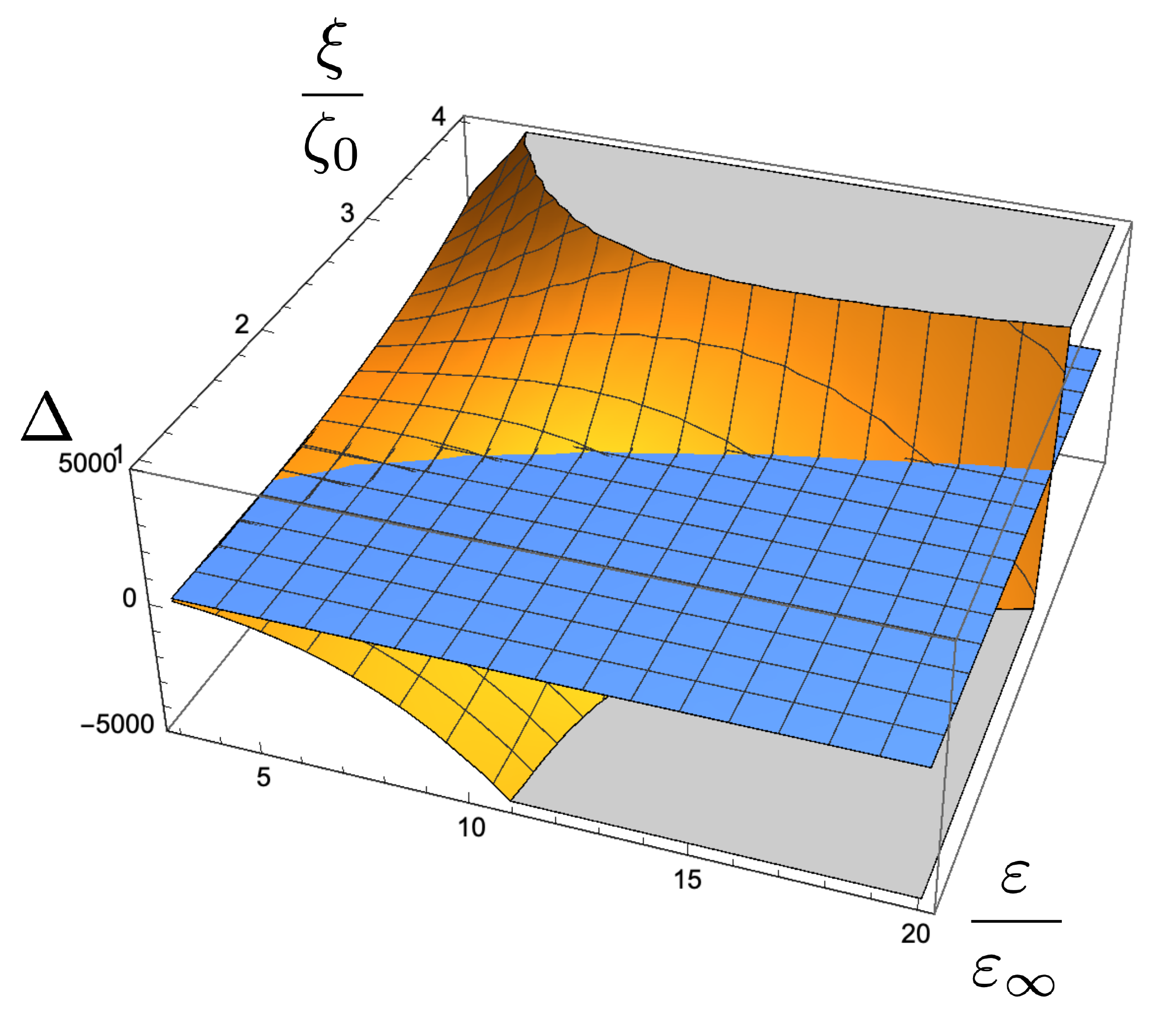}
\end{center}
\caption{The discriminant $\Delta$ for {\sl Model 2} as a function of $\xi/\zeta_0$ and $\varepsilon/\varepsilon_{\infty}$ in relevant parameter ranges; $\kappa_D\widehat{\zeta_0} = 10^{-4}$. For $\Delta > 0$ there are three real roots of the eigenvalue equation, 
while for $\Delta < 0$, one real and two complex conjugate roots arise.}
\end{figure}

Figure 5 displays the numerical solution of the equations of {\sl Model 2} for the case
the plate geometry. The boundary conditions for the fields $\phi^*$ and $\phi^{\dagger} $ 
have been chosen in an antagonistic fashion, which show that both fields behave as separate
entities near the wall. For distances $ z \gg 0$, both fields merge into each other and fulfill the definition of the polarization field of Eq. (\ref{polfield}) in the bulk.
\begin{figure}[t] 
\begin{center}
\includegraphics[width=9cm]{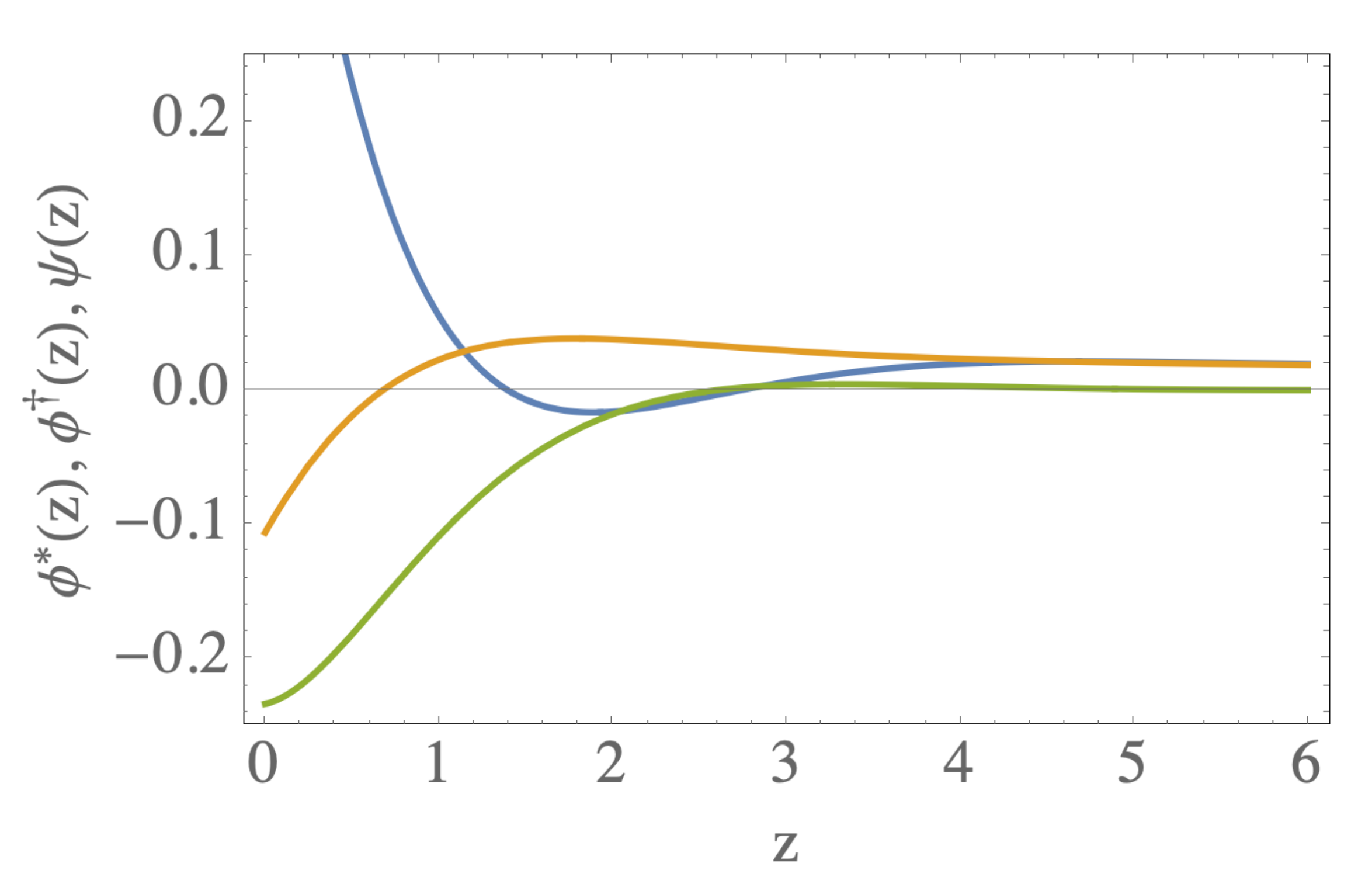}
\end{center}
\caption{Solutions to the equations of {\sl Model 2}. Blue: $\phi^*(z)$, Brown: $\phi^{\dagger}(z)$, Green: auxiliary field $\psi(z)$.
For $z \gg 0$, the solutions of $\phi^*$ and $\phi^{\dagger}$ merge, establishing the
validity of the standard relation between the polarization and electric fields in the bulk.
}    
\label{Figure5}
\end{figure} 

Neglecting ionic screening altogether, {\sl i.e.} $\kappa_D = 0$, the equations for {\sl 
Model 2} can be simplified further and reduced to a single fourth-order equation for the auxiliary field $\psi(z)$ in the form 
\begin{equation}
   \widehat{\zeta_0}^4 \psi^{iv}(z) - {\widehat{\xi}^2}\psi''(z)
+ ({\textstyle\frac{\varepsilon}{\varepsilon_{\infty}}})
\psi(z) = 0\,.
\label{equfourth}
\end{equation}
Apart from the definition of the constants, the basic equation then reduce to the form introduced in Ref. \cite{Monet2021}.  
We note that fourth-order equations can be obtained also by considering quadrupole coupling terms pertinent to the quadrupolarizable solvent models \cite{Slavchov2014a,Slavchov2014b}.

Making the substitution 
$\kappa \longrightarrow \kappa/\hat\zeta_0$, so that the decay length is in units of $\hat\zeta$, the characteristic polynomial for the inverse decay length, Eq.(\ref{equfourth}), can be obtained in the simple form
\begin{eqnarray}
\kappa^4 - \kappa^2~ \left({\textstyle\frac{\widehat{\xi}}{\widehat{\zeta_0}}}\right)^2  + {\textstyle\frac{\varepsilon}{\varepsilon_{\infty}}}
 =  0.
\end{eqnarray}
The solution for the eigenvalues are then of two types: real and complex, depending on the sign of the discriminant 
\begin{equation}
    \Delta =  \left({\textstyle\frac{\widehat{\xi}}{\widehat{\zeta_0}}}\right)^4 - 4 {\textstyle\frac{\varepsilon}{\varepsilon_{\infty}}}. 
\label{disc}
\end{equation} 
which amounts to the identical dependence found in the discriminant $\Delta$ of the
full system of equations. 
In the case of $ \widehat{\xi}^2 > \sqrt{{4 \varepsilon}/{\varepsilon_{\infty}} }~ \widehat{\zeta_0}^2$, the eigenvalues are real and obtained as 
\begin{eqnarray}
\kappa_{1,2,3,4} =  \pm \frac{1}{\sqrt{2}}  
\left(\left({\textstyle\frac{\widehat{\xi}}{\widehat{\zeta_0}}}\right)^2 \pm {\delta}^{1/2}\right)^{1/2},
\end{eqnarray}
while for $ \widehat{\xi}^2 \leq \sqrt{ {4 \varepsilon}/{\varepsilon_{\infty}} }~ \widehat{\zeta_0}^2$, the eigenvalues are complex conjugate pairs and given by
\begin{eqnarray}
\kappa_{1,2,3,4} &=&  \pm \left( \kappa_d \pm i \kappa_o\right) 
\end{eqnarray}
where
\begin{eqnarray}
\kappa_{d,o} = {\textstyle\frac{1}{2} \left({
 \left({\left({\textstyle\frac{\widehat{\xi}}{\widehat{\zeta_0}}}\right)^4+ (-\delta)}\right)^{1/2} \pm  \left({\textstyle\frac{\widehat{\xi}}{\widehat{\zeta_0}}}\right)^2}\right)^{1/2}}.
\end{eqnarray}
This last case is also the one considered in Ref. \cite{Monet2021}. Clearly, even with the fourth-order derivative polarization functional the corresponding characteristic lengths can be real, so that periodic solutions are not universal but depend on the sign of the discriminant Eq. (\ref{disc}). Since $\zeta_0$ is comparable to the size of the molecule and $\xi$ is the macroscopic correlation length, the periodic solution are expected for sufficiently polar media $4 {\varepsilon}/{\varepsilon_{\infty}} \gg 1$.

\subsection{Model 3}

A similar, albeit approximate solution can be found for the case of {\sl Model 3} when one neglects the solvent molecular length $\zeta$ compared to
the ion density-polarization coupling $\alpha$. We then have the coupling matrix
for the composite potential $\Phi_{3}(z) \equiv (\phi^*(z),\phi^{\dagger}(z),\psi(z))$ by the expression
\\
\begin{eqnarray}
{\cal M}_3 = \left(\begin{array}{ccc}
\widehat{\alpha}^{-2} &  -\widehat{\alpha}^{-2}  & \widehat{\alpha}^{-2}  \\
    0        &  0 &   \widehat{\xi}^{-2} \\
        \frac{\left( \frac{\widehat{\xi}}{\widehat{\alpha}}\right)^2 - \frac{\varepsilon}{\varepsilon_{\infty}} (\kappa_D\xi)^2 }{({\textstyle\frac{\varepsilon}{\varepsilon_{\infty}}} -1)\widehat{\alpha}^2}  & - \frac{\left( \frac{\widehat{\xi}}{\widehat{\alpha}}\right)^2  }{({\textstyle\frac{\varepsilon}{\varepsilon_{\infty}}} -1)\widehat{\alpha}^2} & \frac{\left( \frac{\widehat{\xi}}{\widehat{\alpha}}\right)^2 + ({\textstyle\frac{\varepsilon}{\varepsilon_{\infty}}} -1) }{({\textstyle\frac{\varepsilon}{\varepsilon_{\infty}}} -1)\widehat{\alpha}^2}
\end{array}\right)\, . \nonumber\\
~
\end{eqnarray}
The corresponding characteristic polynomial is again of third order and contains four model parameters. Choosing one of them, {\sl e.g.} $\hat\alpha$, and rescaling the roots by $\lambda \longrightarrow \lambda/\hat\alpha^2$ the characteristic polynomial comes out as 
\\
\begin{eqnarray}
   \lambda^3 -  &&
    \lambda^2
    \left[ \frac{\left( \frac{\widehat{\xi}}{\widehat\alpha}\right)^2 + 2 \left(\frac{\varepsilon}{\varepsilon_{\infty}} -1 \right)}{({\textstyle\frac{\varepsilon}{\varepsilon_{\infty}}} -1)} \right]
    + \nonumber\\
    && + ~\lambda 
\frac{\frac{\varepsilon}{\varepsilon_{\infty}} \left( 1 + (\kappa_D \widehat{\xi})^2\right)}{({\textstyle\frac{\varepsilon}{\varepsilon_{\infty}}} -1)}
    - 
    \frac{\frac{\varepsilon}{\varepsilon_{\infty}} (\kappa_D \widehat\alpha)^2}{({\textstyle\frac{\varepsilon}{\varepsilon_{\infty}}} -1)}
      = 0\, .
    \label{eigenvaluesa}
\end{eqnarray}
Comparing this expression with Eq. (\ref{eigenvalues1}) we see that apart from the denominator $({\textstyle\frac{\varepsilon}{\varepsilon_{\infty}}} -1)$ there is an almost complete correspondence between the molecular length $\zeta$ and the {\sl secondary hydration} structural length $\alpha$. The discussion of the discriminant $\Delta$ of the cubic polynomial for {\sl Model 3} therefore shows the same characteristic behaviour as  for {\sl Model 2}. The criterion for the change of the eigenvalue spectrum from 
real to real and complex conjugates is for this model replaced by the expression
\begin{equation}
    {\textstyle\left(\frac{\widehat{\xi}}{\widehat{\alpha}}\right)^2} < 2
    \left[
    1 -\textstyle{\frac{\varepsilon}{\varepsilon_{\infty}}}\left(1 - {\textstyle(1-\frac{\varepsilon_{\infty}}{\varepsilon}})^{1/2} \right)
    \right],
\end{equation}
which in the limit $\varepsilon \gg \varepsilon_{\infty}$ reduces to 
$\xi^2/\widehat{\alpha}^2 < 1$. Since by definition $\widehat{\alpha}^2$ can be negative, in this case the eigenvalues are always real and complex conjugates; in the case $\widehat{\alpha}^2 > 0$, if the correlation length is larger than the ion-density polarization coupling, all eigenvalues are real. The plot of the discriminant for the case $\widehat{\alpha}^2 > 0 $ 
is shown in Figure 6.
\begin{figure}[tbh] 
\begin{center}
\includegraphics[width=8.5cm]{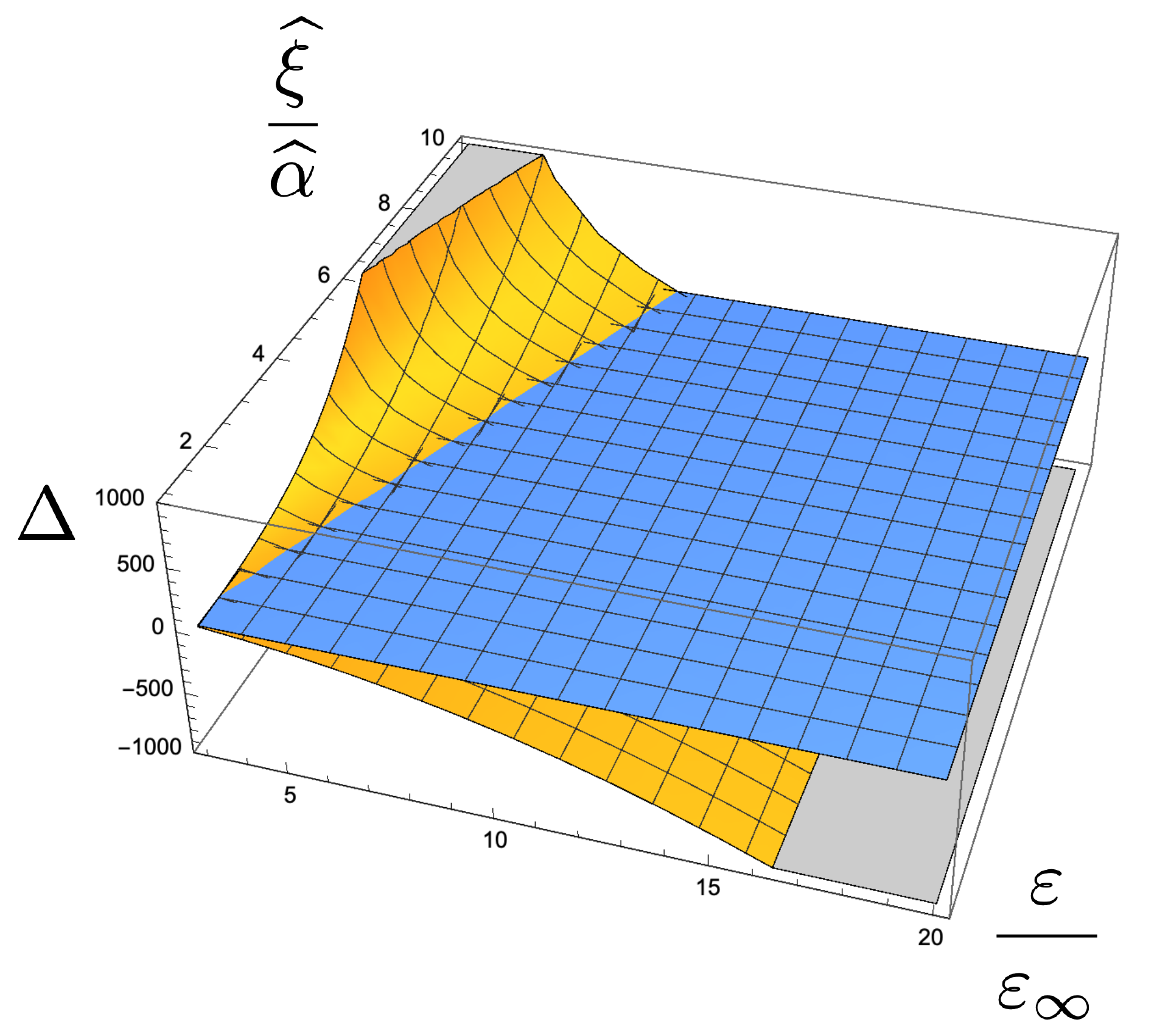}
\end{center}
\caption{The discriminant $\Delta$ for {\sl Model 3} as a function of $\widehat{\xi}/\widehat{\alpha}$ and $\varepsilon/\varepsilon_{\infty}$ in relevant parameter ranges; $\kappa_D\widehat{\zeta} = 10^{-4}$. For $\Delta > 0$ there are three real roots of the eigenvalue equation, 
while for $\Delta < 0$, one real and two complex conjugate roots arise.}
\end{figure}

Profiting further from our insights from {\sl Model 2}, 
we also consider the case of a vanishing Debye-screening length in {\sl Model 3}. 
We then find the expression 
\begin{equation}
\label{mod3}
    \widehat{\zeta_0}^4(\phi^{\dagger})^{iv} - 
    \left(2\widehat{\alpha}^2({\textstyle\frac{\varepsilon}{\varepsilon_{\infty}}} -1)+
    \widehat{\xi}^2\right)(\phi^{\dagger})^{''} + {\textstyle\frac{\varepsilon}{\varepsilon_{\infty}}} \phi^{\dagger} = 0\,.
\end{equation}
Clearly the hydration coupling corresponds to   secondary hydration effects with the square of the effective structural correlation length given by $ 2\widehat{\alpha}^2({\textstyle \frac{\varepsilon}{\varepsilon_{\infty}}} - 1)+ \widehat{\xi}^2 $.
Making the substitution $\kappa \longrightarrow \kappa/\left(({\textstyle\frac{\varepsilon_{\infty}}{\varepsilon}})^{1/4} \widehat{\zeta_0} \right)$,  the characteristic polynomial of Eq.(58) can be obtained in the simple form
\begin{eqnarray}
\kappa^4 - \kappa^2 {\left({\textstyle\frac{\varepsilon_{\infty}}{\varepsilon}} \right)^{1/2} \left(2 \left({\textstyle\frac{\widehat{\alpha}}{\zeta_0}}\right)^2({\textstyle\frac{\varepsilon}{\varepsilon_{\infty}}} -1)+
    \left({\textstyle\frac{\widehat{\xi}}{\widehat{\zeta_0}}}\right)^2\right)} + 1 = 0. \nonumber \\
\end{eqnarray}
Equivalently, one can choose a different rescaling  $\kappa \longrightarrow \kappa/\left(\left({\textstyle\frac{\varepsilon_{\infty}}{\varepsilon}}\right)
    \left(2\widehat{\alpha}^2
    ({\textstyle\frac{\varepsilon}{\varepsilon_{\infty}}} -1) + \widehat{\xi}^2\right)\right)^{1/2}$ which then leads to the characteristic equation 
\begin{equation}
    \left({\textstyle\frac{\varepsilon_{\infty}}{\varepsilon}}\right)^3\widehat{\zeta_0}^4\left(
    2\widehat{\alpha}^2
    ({\textstyle\frac{\varepsilon}{\varepsilon_{\infty}}} -1) + \widehat{\xi}^2\right)^2  \kappa^4 - \kappa^2 + 
    1 = 0\,.
\end{equation}
The case $\left({\textstyle\frac{\varepsilon_{\infty}}{\varepsilon}}\right)^3\widehat{\zeta_0}^4\left(
    2\widehat{\alpha}^2
    ({\textstyle\frac{\varepsilon}{\varepsilon_{\infty}}} -1) + \widehat{\xi}^2\right)^2 > 1/2$ has been discussed in \cite{Bazant2011} in the context of
    ionic liquids but is formally equivalent to the above case. 
The exact profile for the case of a planar charged surface $0 \leq z < \infty$ is given in Appendix D. The corresponding profiles for $\phi^*(z)$ and $\phi^{\dagger}(z)$ are
shown in Figure 7.
\begin{figure}[tbh] 
\begin{center}
\includegraphics[width=9cm]{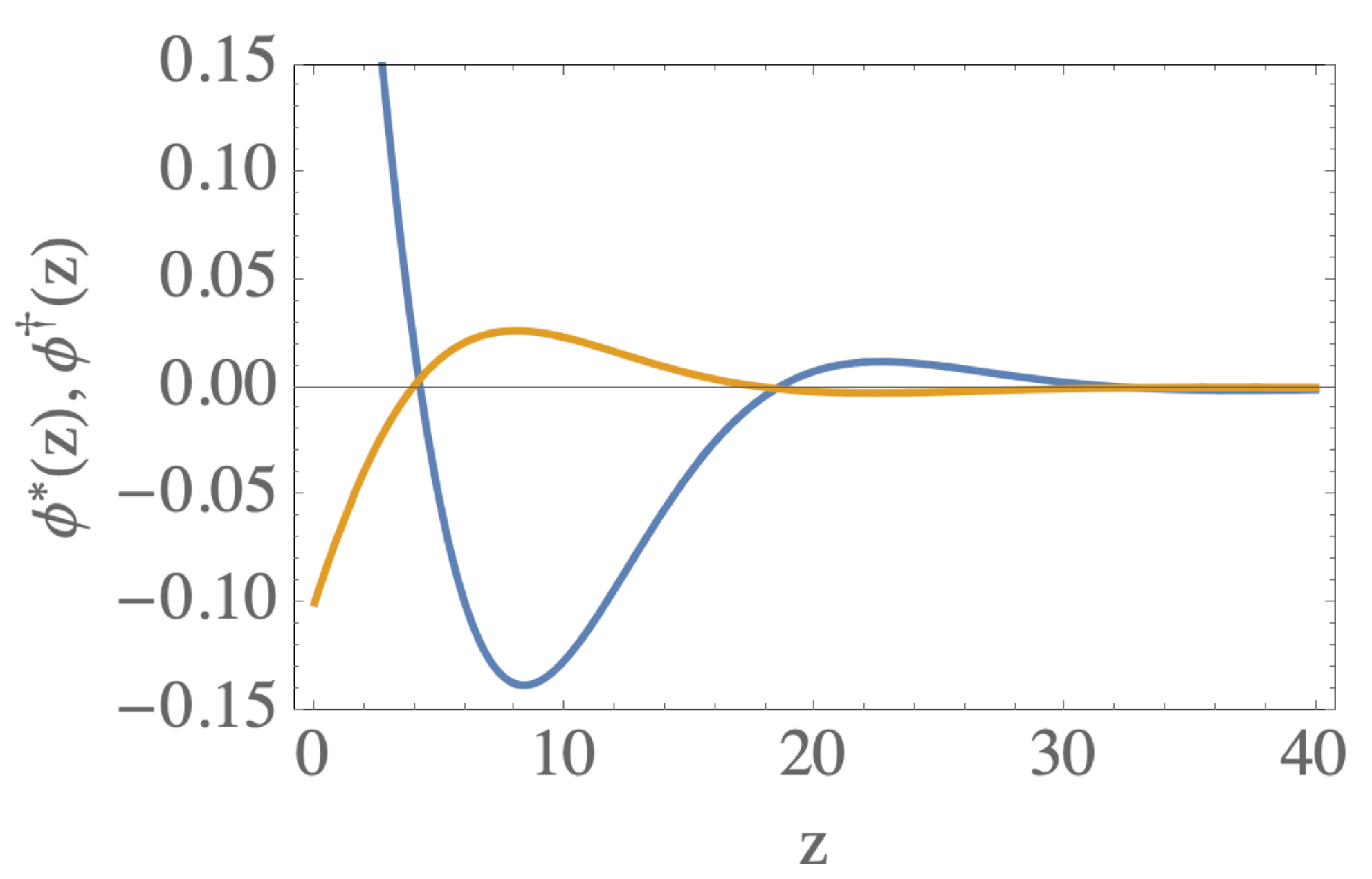}
\end{center}
\caption{Solutions of the equations of the reduced {\sl Model 3}, Eq. (\ref{mod3}); $\phi^*(z)$ (blue line); $\phi^{\dagger}(z)$ (brown line); see Appendix D for the 
analytic expressions for $\phi^*(z)$ and $\phi^{\dagger}(z)$.}  
\end{figure}

\section{Discussion and Conclusions}

In this paper we have derived a field-theoretic description of structured liquid dielectrics, taking the Onsager-Dupuis theory of ice as a motivation. 
Starting from the introduction of {\sl two separate order parameters} for the polarization field of the solvent and the charge density of the solvated ions,
we provide a general formalism to introduce a family of models of different degree of coarse-graining. Considering a harmonic approximation for the
free energy of the polarization field and a dipolar model for the solvent, we derive 
novel nonlinear equations that emerge from the present theory. 

As a first application
we discuss three models defined in the linear limit of the theory, which lead to coupled differential equation systems for the electrostatic and polarization potentials. The generic behaviour of the solutions is determined from a decoupling of these equations. The resulting discriminants of the eigenvalue equations define the solution behaviour as functions of model structural lengths and solvent polarity. Since the model equations allow for independent boundary conditions for the electrostatic and polarization potentials they have a high flexibility in the discussion of physical systems,  which we illustrate for plate and slab geometries. In the case of the ion charge density-solvent polarization coupling our approach allowed us to describe the case of {\sl secondary hydration} effects where the structural solvent correlation length has solvent {\sl as well as} ion contributions.

We believe that, beyond the introduction of our approach and first results in the linear regime, our theory is capable of addressing a wide class of continuum models for polarization phenomena in confined solvents, in particular phenomena pertinent to the nano-confined aqueous systems \cite{Fumagalli2018, Papadopoulou2021}. Further extension of our work to allow for nonlinear polarization functionals in the bulk and the incorporation of surface polarization functionals are also  immediately feasible.

\begin{acknowledgments}
R.P.\ acknowledges the support of the University of Chinese Academy of Sciences and the funding from the NSFC under grant No.\ 12034019.
\end{acknowledgments}

\section{Appendices}

\appendix

\section{Collective description and field theory}

To express the partition function in the form of a field theory, we take the same approach as with anisotropic Coulomb fluids \cite{Podgornik2021}, the outlines of which we briefly reiterate below.

We write the total energy with two collective variables, the polarization and the charge density, Eqs. (\ref{cole2}) and (\ref{cole3}), by introducing the decomposition of unity
\begin{eqnarray}
1 \equiv  \int {\cal D}[{\cal \bf P}_{i}({\bf x})] ~\Pi_{{\bf x}} \delta({\cal \bf P}_{i}({\bf x}) - \hat{\cal \bf P}_{i}({\bf x})) \times \nonumber\\
\int {\cal D}[\rho({\bf x})] ~\Pi_{{\bf x}} \delta(\rho({\bf x}) - \hat{\rho}({\bf x}))
\end{eqnarray}
together with the functional integral representation of the delta functions  in terms of their  respective auxiliary fields {${\cal E}_i(x)$, $\phi(x)$}:
\begin{eqnarray}
&& \Pi_{{\bf x}} \delta({\cal \bf P}_{i}({\bf x}) - \hat{\cal \bf P}_{i}({\bf x})) \longrightarrow \int {\cal D}[{\cal E}_{i}({\bf x})]  ~e^{- i {\cal E}_i({\bf x}) ({\cal \bf P}_{i}({\bf x}) - \hat{\cal \bf P}_{i}({\bf x}))} \nonumber\\
&&  \Pi_{{\bf x}} \delta(\rho({\bf x}) - \hat{\rho}({\bf x})) \longrightarrow \int {\cal D}[\phi({\bf x})]~e^{- i \phi({\bf x}) (\rho({\bf x}) - \hat{\rho}({\bf x}))}.
\end{eqnarray}
Notably, we introduce the microscopic polarization as a separate order parameter, independently of the ion charge density. This will eventually lead to a two-order parameter representation of the partition function.

Using these relations in the definition of the grand canonical partition function 
\begin{eqnarray}
{\Xi} &\equiv& \sum_{N} \frac{\lambda^N}{N!} \sum_{N_+}\sum_{N_-} \frac{\lambda_{(+)}^{N_+} \lambda_{(-)}^{N_-}}{N_+!N_-!} \nonumber\\
&& \int {\cal D}[{\bf x}_N] {\cal D}[{\bf n}_N]\int {\cal D}[{\bf x}_{N_{\pm}}] ~ e^{ -\beta {\cal H}[{\bf x}_N, {\bf x}_{N_{\pm}}]},
\end{eqnarray}
with $\lambda$ and $\lambda_{(\pm )}$ as the absolute activities of the dipolar and ionic components,  the introduction of collective variables ${\cal \bf P}_{i}({\bf x})$ and $\rho({\bf x})$ leads to the following field-representation of the partition function
\begin{eqnarray}
{\Xi} &=&\int {\cal D}[{\cal \bf P}_{i}({\bf x})]{\cal D}[{\cal E}_i({\bf x})]  \int {\cal D}[\rho({\bf x})]{\cal D}[\phi({\bf x})]~ \times \nonumber\\
&& e^{-\beta \tilde{\cal H}[{\cal \bf P}_{i}({\bf x}),~\rho({\bf x}); {\cal E}_i({\bf x}), \phi({\bf x})]  - V[{\cal E}_i({\bf x}), \phi({\bf x})]}. 
\end{eqnarray}
The field action above decouples into two terms, of which the first one $\beta \tilde{\cal H}[{\cal \bf P}_{i}({\bf x}),~\rho({\bf x}); {\cal E}_i({\bf x}), \phi({\bf x})]$ 
depends on the collective order parameters ${\cal \bf P}_{i}({\bf x})$ and $\rho({\bf x})$ as well as the auxiliary fields 
\begin{eqnarray}
&& \beta \tilde{\cal H}[{\cal \bf P}_{i}({\bf x}),~\rho({\bf x}); {\cal E}_i({\bf x}), \phi({\bf x})] =  \nonumber\\
&& ~~~~~{\textstyle\frac12} \int\!\!\int_V d{\bf x} d{\bf x}' ~ {\cal \bf P}_{i}({\bf x})\tilde u_{ij}({\bf x} -{\bf x}') {\cal \bf P}_{j}({\bf x}') + \nonumber\\
&& ~~~~~ {\textstyle\frac12} \int\!\!\int_V d{\bf x} d{\bf x}' ~ {\rho}({\bf x})u({\bf x} -{\bf x}') {\rho}({\bf x}') + \nonumber\\
&& ~~~~~  i \int_V d{\bf x}~ {\cal \bf P}_{i}({\bf x}) {\cal E}_{i}({\bf x}) + i \int_V d{\bf x} ~\rho({\bf x})\phi({\bf x}),
\end{eqnarray}
and is a sum of the interaction Hamiltonian for non-electrostatic  polarization interactions and Coulomb interactions between mobile and bound charges. 

The second component, the effective field interaction potential,  corresponding to the integration over the configurational variables ${\bf x}_N$, ${\bf n}_N$ and ${\bf x}_{N_{\pm}}$ of the solvent dipoles and electrolyte ions, depends only on the auxiliary fields ${\cal E}_i({\bf x})$ and $\phi({\bf x})$ as
\begin{widetext}
\begin{eqnarray}
V[{\cal E}_i({\bf x}), \phi({\bf x})] &\equiv&  \log{\Big(\sum_{N}\ \sum_{N^+}{\sum_{N^-}} \frac{\lambda^N}{N!}  \frac{\lambda_{(+)}^{N^+} \lambda_{(-)}^{N^-}}{N^+!N^-!}  \int {\cal D}[{\bf x}_N] {\cal D}[{\bf n}_N]~\int {\cal D}[{\bf x}_{N_{\pm}}] ~ e^{ -\beta \tilde{\cal H}^*[ {\cal E}_{i}({\bf x}_i), \phi({\bf x}_i)]}\Big)}  \nonumber\\
& =  & \lambda \int_V d^3{\bf x} \int_{\Omega}d{\bf n} ~e^{- i p ~\!{\bf n}_i ~\! \big( {\cal E}_i({\bf x})-\bnabla_i \phi({\bf x}) \big)} + 2 \lambda_s \int_V d^3{\bf x} \cos{\beta e \phi({\bf x})} \nonumber\\
& = & \int_V d^3{\bf x} ~\Big( \lambda \int_{\Omega}\sin{\theta} d\theta ~e^{  - i p\cos{\theta}  ~\vert {\cal E}_i({\bf x})-\bnabla_i \phi({\bf x}) \vert} + 2 \lambda_s  \cos{\beta e \phi({\bf x})}\Big) \nonumber\\
& = &  \lambda \int_V d^3{\bf x} ~\Big( \frac{  \sin{p \vert {\cal E}_i({\bf x})-\bnabla_i \phi({\bf x}) \vert}}{p \vert {\cal E}_i({\bf x})-\bnabla_i \phi({\bf x}) \vert}+ 2 \tilde\lambda_s  \cos{\beta e \phi({\bf x})}\Big) = \int_V d^3{\bf r}~ v\left({\cal E}_i({\bf x}), \phi({\bf x})\right), \nonumber\\
~
\label{Xi-0}
\end{eqnarray}
\end{widetext}
where the single-particle configurational Hamiltonian is
\begin{eqnarray}
\beta \tilde{\cal H}^*[{\cal E}_{i}({\bf x}_i), \phi({\bf x}_i)] 
&=&  - i~ p \sum_{N}   {\bf n}_i ~\! \big( {\cal E}_i({\bf x}_i)-\bnabla_i \phi({\bf x}_i) \big) - \nonumber\\
&& - i \sum_{N_+} ~\phi({\bf x}_i) +  i  \sum_{N_-} ~\phi({\bf x}_i). 
~
\end{eqnarray}
We have rescaled all the interaction energies in terms of the thermal energy and assumed that the salt ion activities are the same, $\lambda_{(+)} = \lambda_{(-)} = \lambda_{s}$  \cite{Abrashkin2007,Levy2013}. Note that above we also redefined $\lambda_s/\lambda \longrightarrow \tilde\lambda_s$. 

Eq. (\ref{Xi-0}) describes an ideal van't Hoff gas of ions and dipoles, corresponding to the  dipolar model (DM). 
This expression differs from the standard Coulomb fluid field-theoretical representation of the partition function \cite{Edwards1962,Podgornik1989} by the presence of two additional independent fields ${\cal \bf P}_{i}({\bf x})$ and ${\cal E}_i({\bf x})$. 
The saddle-point evaluation of the partition function is taken with respect to the auxiliary fields ${\cal E}_i({\bf x}) \longrightarrow i{\cal E}^*_i({\bf x})$ and $\phi({\bf x}) \longrightarrow i \phi^*({\bf x}) $. The partition function is then approximated by 
\begin{eqnarray}
- \ln{{\Xi}} \longrightarrow  \beta {\cal H}[{\cal \bf P}_i({\bf x}), \rho({\bf x}); i{\cal E}^*_i({\bf x}), i\phi^*({\bf x})],
\end{eqnarray}
where the effective-field Hamiltonian is obtained from the partial trace over coordinate and orientation degrees of freedom for the ions and dipoles in the partition function interacting {\sl via} the interaction energy functional Eq. (\ref{Hamiltonian}). 

The expression for the effective field interaction potential $V^*[{\cal E}_i({\bf x}), \phi^*({\bf x})] = V[i{\cal E}^*_i({\bf x}), i \phi^*({\bf x})]$ depends crucially on the model used for the dipolar fluid, see main text.  
The saddle-point form of the DM  effective field interaction potential $V^*[i{\cal E}^*_i({\bf x}), i \phi^*({\bf x})]$ can be finally written explicitly in the form of Eq. (\ref{Xi-0}) but with $\sin{\beta p \vert {\cal E}^*({\bf x})-\bnabla \phi^*({\bf x}) \vert} \longrightarrow \sinh{\beta p \vert {\cal E}^*({\bf x})-\bnabla \phi^*({\bf x}) \vert}$ and $\cos{\beta e \phi^*({\bf x})} \longrightarrow \cosh{\beta e \phi^*({\bf x})}$.

\section{Hydration shell coupling and the derivation of {\sl Model 3}}

Since it couples the ion density and polarization, the microscopic hydration energy of {\sl Model 3} is actually of the form
\begin{eqnarray}
{{\cal H}_{HY}} &=& {\textstyle\frac12} \int\!\!\int_V d{\bf x} d{\bf x}' ~\hat{\rho}_{(\pm )}({\bf x})  \tilde{u}({\bf x}\!-\!{\bf x}') \bnabla\!\cdot\!\hat{\cal \bf P}({\bf x}'),
\end{eqnarray}
where the potential $\tilde u({\bf x} -{\bf x}')$ is a {short-range, non-electrostatic potential} describing the hydration shell interactions. We assumed that the hydration polarization for anions and cations is - apart from the direction - the same. This could be of course elaborated further.  Notice, however, that the hydration energy can be recast as
\begin{eqnarray}
{{\cal H}_{HY}} &=& {\textstyle\frac12}\!\! \int\!\!\int_V\!\!d{\bf x} d{\bf x}'\!\!\left(  {\hat\rho}({\bf x}) - \bnabla\!\cdot\!\hat{\cal \bf P}({\bf x}) \right) \tilde{u}({\bf x}\!-\!{\bf x}') \bnabla\!\cdot\!\hat{\cal \bf P}({\bf x}'), \nonumber\\
~
\label{hamilt3a}
\end{eqnarray}
so that the last term would just rescale the potential $\tilde u_{ij}({\bf x} -{\bf x}')$ in Eq. \ref{hamilt0} 
\begin{eqnarray}
\tilde u_{ij}({\bf x} -{\bf x}') \longrightarrow \tilde u_{ij}({\bf x} -{\bf x}') + \tilde{u}({\bf x} -{\bf x}')\bnabla_i\bnabla'_j,  
\end{eqnarray}
and if we take the non-electrostatic potential to the second order in derivatives this simply means a rescaling of the corresponding  constants, which is irrelevant. The hydration energy can thus be simply taken as  
  \begin{equation}
      \int\!\!\int_V d{\bf x} d{\bf x}' ~ \hat{\rho}({\bf x})\tilde u({\bf x} -{\bf x}') \bnabla \cdot{\hat{\cal \bf P}}({\bf x}'), 
      \label{equcoupling}
 \end{equation}
 which is what we use in the main text, Eq. (\ref{hamilt3}).  If the hydration shell of anions and cations differ, the proper {\sl Ansatz} would then be
  \begin{eqnarray}
     && \int\!\!\int_V d{\bf x} d{\bf x}' ~ \hat{\rho}_{(+)}({\bf x})\tilde u_{+}({\bf x} -{\bf x}') \bnabla \cdot{\cal \bf \hat P}({\bf x}') + \nonumber\\
     && + \int\!\!\int_V d{\bf x} d{\bf x}' ~ \hat{\rho}_{(-)}({\bf x})\tilde u_{-}({\bf x} -{\bf x}') \bnabla \cdot{\cal \bf \hat P}({\bf x}'), 
 \end{eqnarray} 
 where the $\tilde u$'s are now the non-electrostatic hydration shell potentials, different for different types of ions. 

In the case of the mean-field approximation, the {\sl Model 3} saddle-point equations are reduced to  
\begin{eqnarray}
\phi^*({\bf x}) = \int_V d{\bf x}' ~u({\bf x} -{\bf x}') {\rho}({\bf x}') + \\
\int_V d{\bf x}' ~\tilde u({\bf x} -{\bf x}') \bnabla \cdot{{\cal \bf P}}({\bf x}') \nonumber 
\label{pol12}
\end{eqnarray}
and
\begin{eqnarray}
{\cal E}^*({\bf x}) =  - \alpha \bnabla {\rho}({\bf x}) + u_P(0){\bf P}({\bf x}) -
{\cal F}(\bnabla \cdot {\bf P}({\bf x}))
\label{sp-nonlocal-linear11}
\end{eqnarray}
where ${\cal F}$ is considered up to fourth order. With
\begin{eqnarray}
\tilde u({\bf x} -{\bf x}') = \alpha ~\delta({\bf x} -{\bf x}') +  \dots 
\label{shortrange1a}
\end{eqnarray}
the expression for $\phi^*({\bf x})$ simplifies to 
\begin{eqnarray}
\phi^*({\bf x}) = \int_V d{\bf x}' ~u({\bf x} -{\bf x}') {\rho}({\bf x}') + \alpha \bnabla \cdot{{\cal \bf P}}({\bf x})\, .
\label{pol121}
\end{eqnarray}
Applying the Laplacian on the last equation yields
\begin{eqnarray}
 {\rho}({\bf x}) = \alpha  \varepsilon_{\infty}\varepsilon_0 \nabla^2 \bnabla \cdot{{\cal \bf P}}({\bf x}) - \varepsilon_{\infty}\varepsilon_0 \nabla^2\phi^*({\bf x}).
\label{pol121}
\end{eqnarray}
Combining everything together into the final set of saddle-point equations for the ion density polarization coupling case we remain with
\begin{equation}
{\cal E}^*({\bf x}) = u_P(0) - {\cal F}(\bnabla \cdot {\bf P}({\bf x})) 
\end{equation}
\begin{eqnarray} 
- \alpha^2 \varepsilon_{\infty}\varepsilon_0 \nabla^2\bnabla\left( \bnabla\cdot {\bf P}({\bf x})\right) + \alpha \varepsilon_{\infty}\varepsilon_0 \nabla^2 (\bnabla\phi^*({\bf x})) \, . \nonumber 
\label{sp-nonlocal-linear11ab}
\end{eqnarray}
The $\alpha$-coupling therefore modifies the fourth-order coefficient of the expansion of 
the non-electrostatic potential.
\\

\section{Model parametrization} 

For numerical calculations of the solutions shown in the paper we employ  
parameters chosen according to the ranges indicated in Table I. 
\begin{table}[ht]
    \centering
    \begin{tabular}{|c|c|c|}
    \hline
    Symbol & Meaning & Value \\
    \hline 
        $1/\kappa_D$ & Debye length & $\sim$ 10 nm\\
        $\xi$  & bulk water correlation length & 1-10 nm\\
        $\zeta$ & water molecular size & 0.3 nm\\
        $\varepsilon$ & bulk dielectric constant & 5-80 \\
        $\varepsilon_{\infty}$ & dielectric constant first dispersion & 4-10\\
        $\alpha$ & ion density-polarization coupling & $\pm $ 0.5 - 5 nm \\
    \hline
    \end{tabular}
    \caption{Model parameters}
    \label{tab:my_label}
\end{table}
The coupling parameter of the structural interactions, $u_P(0)\varepsilon_0 $, can be obtained from Eq. (26) in the main text, slightly rewritten in order to make the dimensional dependencies explicit. It then reads as
\begin{equation}
     {\textstyle\left(\frac{\varepsilon}{\varepsilon_0} - \frac{\varepsilon_{\infty}}{\varepsilon_0}\right)}
     =  
     \frac{{\textstyle\frac{\lambda p^2}{3 \varepsilon_0}}}{1 + u_P(0)\varepsilon_0{\textstyle\frac{\lambda p^2}{3\varepsilon_0}}},
 \end{equation}
where $\lambda$ is the fugacity and $p$ the dipole moment in the dipolar model. The form of this relation would remain unchanged if one considers the dipolar Langevin model, but the values of the constants would have a different meaning. We consider these differences as irrelevant at this stage, since we are mainly interested here in qualitative features of the
theory. By dimensional analysis, one has
\begin{equation}
    [\lambda] = J^{-1} m^{-3}\,,\,\,\, 
    [p] = C m\,,\,\,\, 
    [\varepsilon_0] = \frac{C^2}{Jm}\, .
\end{equation}
In order to connect the value for $u_P(0)\varepsilon_0$ with a measurable macroscopic property, we invoke an additional relation
\begin{eqnarray}
    { 1 + \frac{\lambda p^2}{3 \varepsilon_0} = \frac{\varepsilon}{\varepsilon_0}} \\
    ~\nonumber
\end{eqnarray} 
which has been used \cite{Abrashkin2007}. 
For the standard values for water, $\varepsilon = 80$, $\varepsilon_{\infty} = 4$, 
we have $u_P(0)\varepsilon_0 = 5 \cdot 10^{-4}$. The hatted parameters can then be computed from the bare lengths by multiplication with a factor $\approx 0.038$. 

\section{The analytic solution for the reduced {\sl Model 3} in the plate geometry}

The analytic solution for the reduced {\sl Model 3} in the slab geometry can be obtained by making recourse to the SI of \cite{Bazant2011}. Defining
\begin{equation}
    \delta_c^2 \equiv \left({\textstyle\frac{\varepsilon_{\infty}}{\varepsilon}}\right)^3\widehat{\zeta_0}^4\left(
    2\widehat{\alpha}^2
    ({\textstyle\frac{\varepsilon}{\varepsilon_{\infty}}} -1) + \widehat{\xi}^2\right)^2 
\end{equation}
one has, for $\delta_c > 1/2$, the exact solution in the slab geometry given by
\begin{equation}
    \phi^{\dagger}(z) = P_c e^{-\kappa_1 z}[\cos(\kappa_2z) + A\sin(\kappa_2 z)]
\end{equation}
with
\begin{equation}
    \kappa_1 = \frac{\sqrt{2\delta_c + 1}}{2\delta_c}\,,\,\, 
    \kappa_2 = \frac{\sqrt{2\delta_c - 1}}{2\delta_c}
\end{equation}
and
\begin{equation}
    A = - \frac{\sqrt{2\delta_c + 1}(\delta_c - 1)}{\sqrt{2\delta_c - 1}(\delta_c + 1)}\, .
\end{equation}
From this solution, $\phi^*(z)$ follows via the expression
\begin{equation} \label{phi*-phi-dagger2}
    \phi^*(z) = -\left(\frac{\varepsilon}{\varepsilon_{\infty}} - 1\right)
    \left(1 - \widehat{\alpha}^2\gamma\frac{d^2}{dz^2}\right)\phi^{\dagger}(z)\, 
\end{equation}
with
\begin{equation}
    \gamma \equiv {\textstyle\left(2\widehat{\alpha}^2\left(\frac{\varepsilon}{\varepsilon_{\infty}} - 1\right) + \widehat{\zeta}^2\right)^{-1}}\, .
\end{equation}

\end{document}